\documentclass{vgtc}                          

\graphicspath{{figures/}{pictures/}{images/}{./}} 

\usepackage{times}                     

\usepackage{tabu}                      
\usepackage{booktabs}                  
\usepackage{lipsum}                    
\usepackage{mwe}                       
\usepackage{siunitx}
\usepackage{enumitem}

\usepackage{graphicx}
\usepackage{multirow}
\usepackage{array}
\usepackage{tabularx}
\usepackage{arydshln}

\usepackage[acronym,shortcuts]{glossaries}
\newacronym{BLV}{BLV}{blind and low vision}
\newacronym{BCI}{BCI}{brain-computer interface}
\newacronym{HCI}{HCI}{human-computer interaction}
\newacronym{HMD}{HMD}{head-mounted display}
\newacronym{OandM}{O\&M}{orientation and mobility}
\newacronym{AR}{AR}{augmented reality}
\newacronym{VR}{VR}{virtual reality}
\newacronym{XR}{XR}{extended reality}
\newacronym{FOV}{FOV}{field of view}

\usepackage{mathptmx}                  

\onlineid{0}

\vgtccategory{Position Paper}

\vgtcinsertpkg

\title{Bionic Vision as Neuroadaptive XR: \\ Closed-Loop Perceptual Interfaces for Neurotechnology}

\author{Michael Beyeler\thanks{e-mail: mbeyeler@ucsb.edu} \\
    \scriptsize University of California, Santa Barbara}

\teaser{
  \centering
  \includegraphics[width=\linewidth]{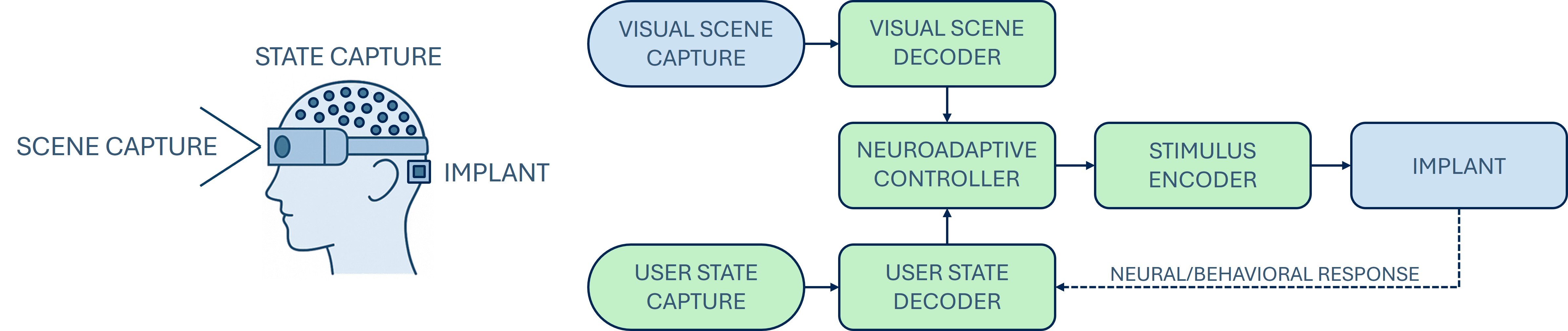}
  \caption{
    Closed‑loop architecture for neuroadaptive XR in bionic vision.
    Blue boxes indicate components found in current systems; green boxes represent proposed modules.
    Visual input (e.g., from RGB-D or stereo cameras) and user-state signals (e.g., pupil dynamics, IMU, implant telemetry; optional EEG/fNIRS) are processed by dedicated decoders.
    A neuroadaptive controller prioritizes scene content based on cognitive state and task demands, guiding a stimulus encoder to produce targeted neural stimulation.
    Feedback from neural and behavioral responses enables adaptive, co-constructed perception.}
  \label{fig:teaser}
}

\abstract{
    Visual neuroprostheses are commonly framed as technologies to restore natural sight to people who are blind. 
    In practice, they create a novel mode of perception shaped by sparse, distorted, and unstable input. They resemble early extended reality (XR) headsets more than natural vision, streaming video from a head-mounted camera to a neural ``display'' with under 1000 pixels, limited field of view, low refresh rates, and nonlinear spatial mappings. 
    No amount of resolution alone will make this experience natural.
    This paper proposes a reframing: bionic vision as neuroadaptive XR.
    Rather than replicating natural sight, the goal is to co-adapt brain and device through a bidirectional interface that responds to neural constraints, behavioral goals, and cognitive state.
    By comparing traditional XR, current implants, and proposed neuroadaptive systems, it introduces a new design space for inclusive, brain-aware computing. 
    It concludes with research provocations spanning encoding, evaluation, learning, and ethics, and invites the XR community to help shape the future of sensory augmentation.
} 

\keywords{bionic vision, brain-computer interface, assistive technology, blindness}

\begin{document}


\firstsection{Introduction}

\maketitle

\Ac{XR} technologies have transformed human-computer interaction through immersive overlays, spatialized cues, and adaptive rendering \cite{adriana_cardenas-robledo_extended_2022,vasarainen_systematic_2021,davari_towards_2024}.
Yet these systems presume intact vision, effectively excluding over 40 million people with incurable blindness \cite{kasowski_systematic_2023}.

Visual neuroprostheses seek to restore a form of vision to people who are blind by electrically stimulating neurons in the visual pathway \cite{palanker_restoration_2025,fernandez_development_2018}.
A head-mounted camera captures the scene, which is processed and translated into electrical pulses delivered to an implanted electrode array (Fig.~\ref{fig:teaser}, \emph{left}).
These pulses evoke flashes of light (\emph{phosphenes}) that users gradually learn to interpret as visual information \cite{erickson-davis_what_2021}.

But the experience is far from natural.
Phosphenes are often distorted, unstable, and idiosyncratic \cite{beyeler_learning_2017,fine_pulse_2015}.
Most systems offer fewer than 1000 electrodes, a field of view under \SI{20}{\degree}, low refresh rates ($6$–\SI{20}{\hertz}), and no encoding of color or depth.
Visual ``pixels" do not sum linearly into contours or forms \cite{barry_video-mode_2020,hou_axonal_2024}, and percepts frequently fade or flicker over time \cite{wilke_fading_2011,perez_fornos_temporal_2012}.

Though often cast as efforts to restore sight \cite{palanker_restoration_2025,roska_restoring_2018}, this framing has long failed to reflect the perceptual reality. 
Bionic vision more closely resembles a constrained form of \ac{XR}, in which a head-mounted sensor streams simplified input to a neural ``display'' with severe limitations in resolution, latency, and perceptual stability \cite{erickson-davis_what_2021,beyeler_towards_2022}.
Efforts to increase the number of implanted electrodes have not resolved these issues---natural vision remains out of reach.

This paper proposes a reframing.
Rather than pursuing a (degraded) imitation of natural sight, bionic vision might be better understood as a form of neuroadaptive XR: a perceptual interface that forgoes visual fidelity in favor of delivering sparse, personalized cues shaped (at its full potential) by user intent, behavioral context, and cognitive state (Fig.~\ref{fig:teaser}B).
Unlike traditional brain-computer interfaces, its goal is not communication or control but \emph{perception}; that is, constructing a usable visual experience under extreme constraints.
This perspective shifts the design objective from visual realism to perceptual utility, inviting a broader research agenda grounded in interaction design, brain-aware computing, and the science of adaptive interfaces.

\begin{table*}[!th]
    \centering
    \scriptsize
    \renewcommand{\arraystretch}{1.3}
    \begin{tabularx}{\linewidth}{l lXXX}
         & \textbf{Component} & \textbf{Traditional XR} & \textbf{Bionic Vision (Current)} & \textbf{Neuroadaptive XR (Proposed)} \\
        \hline
        \multirow{3}{*}{\raisebox{-0.5\totalheight}{\rotatebox{90}{Sensing}}} 
        & Capture Modalities & RGB/-D*, stereo*, thermal* & RGB & RGB-D, stereo*, thermal* \cite{sadeghi_benefits_2024} \\
        & Head Tracking & IMU + SLAM & None & IMU + SLAM \\
        & Eye Tracking & Foveation*, intent inference* & None & Pupil size, eye movements \cite{caspi_eye_2021} \\
        \cdashline{2-5}

        \multirow{3}{*}{\raisebox{-.95\totalheight}{\rotatebox{90}{Scene Proc.}}} 
        & Scene Geometry & Dense 3D maps, SLAM surfaces & None & Layout, spatial primitives \cite{sanchez-garcia_semantic_2020,rasla_relative_2022} \\
        & Semantic Interpretation & Object/region segmentation; scene graphs & None & Object affordances, task-relevant categories \cite{beyeler_towards_2022,han_deep_2021} \\
        & Feature Prioritization & Gaze/context filters & None & Intent-aware filtering \\
        \cdashline{2-5}

        \multirow{4}{*}{\raisebox{-.75\totalheight}{\rotatebox{90}{Rendering}}} 
        & Display Substrate & OLED, microLED, optics & Retina/visual cortex & Retina/visual cortex \cite{palanker_restoration_2025,fernandez_development_2018} \\
        & Spatial Layout & High-res, rectangular display & Low-res ($<1000$ ``pixels''), distorted & Retinotopic mapping \cite{striem-amit_functional_2015} \\
        & Color Encoding & RGB, HDR*, dynamic range* & Grayscale & Symbolic or high-contrast mappings \\
        & Luminance Control & Brightness/contrast & Amplitude modulation & Adaptive brightness modulation \\
        \cdashline{2-5}

        \multirow{4}{*}{\raisebox{-\totalheight}{\rotatebox{90}{Control \& Feedback}}} 
        & User State Decoder & Gaze-aware UIs*, affect models* & None & Pupil, IMU, implant telemetry; fNIRS/EEG* (optional) \\
        & Controller & State-aware rendering, personalization & Manual gain control & Constraint-aware filter for salience, safety, comfort \\
        & Adaptation Strategy & Rule-based or learned UI tuning & Manual parameter tuning & Continual learning \cite{eckmiller_learning_2009} \\
        & Feedback Signals & Haptics*, biosignals* & None & Real-time biosignals, implant feedback, percept quality \\
        \cdashline{2-5}

        \multirow{3}{*}{\raisebox{-1.5\totalheight}{\rotatebox{90}{Output}}} 
        & Output Modalities & Visual (screen), spatial audio*, haptics* & Visual (phosphenes) & Visual (phosphenes), spatial audio*, haptics* \\
        & Spatial Characteristics & High-res, full field of view & Sparse; rastered electrode patterns & Spatial filtering \\
        & Temporal Characteristics & $>\SI{60}{\hertz}$, low latency & $6\text{–}\SI{20}{\hertz}$; fading, persistence & Adaptive update rate; multimodal integration \\
        \cdashline{2-5}

        \multirow{3}{*}{\raisebox{-\totalheight}{\rotatebox{90}{Evaluation}}} 
        & Objective Metrics & Task time, accuracy, presence & Acuity, visual function & Task-level benchmarks \cite{beyeler_towards_2022} \\
        & Subjective Metrics & Usability, immersion & FLORA \cite{geruschat_flora_2015} & Trust, agency, load, usability \\
        & UX Design Approach & Iterative user testing & Engineer-driven & Participatory design, blind-user co-development \\
    \end{tabularx}
    \caption{Comparison of XR pipeline stages. *Optional features depending on user profile, device configuration, or study design.}
    \label{tab:pipeline}
\end{table*}

\section{Bionic Vision as Neuroadaptive XR}

Visual prostheses enable a new mode of perception mediated by neural stimulation, constrained sensing, and active cognitive interpretation.
This calls for a reimagining of each stage in the XR pipeline through the lens of \emph{neuroadaptive XR} (Table~\ref{tab:pipeline}).

\subsection{Sensing}

Modern XR systems draw on a rich stream of multisensory input (e.g., RGB-D, stereo, inertial tracking, eye tracking) to model both the environment and the user.
In contrast, current visual prostheses typically rely on a single RGB camera mounted on a pair of glasses. This creates a perceptual bottleneck: users must suppress natural eye movements and instead scan the scene with head motion, often yielding a fragmented and cognitively demanding experience \cite{erickson-davis_what_2021}.

Still, the sensing landscape is advancing.
Prototypes have begun incorporating gaze tracking for retinotopically aligned displays \cite{caspi_eye_2021}, thermal cameras to visualize warm bodies beyond the visible spectrum \cite{sadeghi_benefits_2024}, and depth sensors to aid obstacle avoidance \cite{mccarthy_importance_2014,rasla_relative_2022}.
Together, these modalities offer a foundation for context-aware, sensor-rich prostheses capable of supporting more intuitive and adaptive interaction.

\subsection{Scene Processing}

Traditional XR platforms construct dense 3D maps and semantic graphs to support navigation and interaction. 
For visual prostheses, such high-fidelity representations are computationally burdensome and perceptually mismatched. 
Instead, the emphasis shifts to lightweight abstractions: spatial primitives (e.g., walls, corridors, obstacles) and task-relevant affordances \cite{beyeler_towards_2022}.

This mirrors strategies from low-vision accessibility tools, which filter and prioritize scene elements based on functional relevance. Prior work has explored semantic segmentation in prosthetic vision to emphasize salient objects \cite{han_deep_2021,horne_semantic_2016}, but opportunities remain to integrate this with spatial understanding and task context.

\subsection{Rendering}

Unlike XR systems that render to dense, high-resolution displays, bionic vision delivers information through sparse, implanted electrode arrays.
These arrays target the retina or cortex and produce phosphenes via amplitude-modulated biphasic pulse trains. 
Stimulation patterns are retinotopically mapped but shaped by complex, nonlinear factors: cortical magnification, individual neural variability, and reorganization due to vision loss \cite{striem-amit_functional_2015,beyeler_learning_2017}.
To optimize this electrode-to-percept transformation, recent advances employ deep learning  \cite{granley_hybrid_2022} and human-in-the-loop calibration \cite{granley_human---loop_2023}.
Yet most implants still produce only achromatic phosphenes and explore only a narrow region of the full stimulus space.

A more radical shift would be to abandon photorealism entirely. Symbolic encodings (i.e., abstract visual motifs tailored to specific tasks) may offer greater interpretability and reduce cognitive load under extreme perceptual constraints.

\subsection{User Input and Adaptation}

XR systems increasingly adapt to users via multimodal input: eye, hand, and voice control; gaze-contingent rendering \cite{plopski_eye_2022}; and physiological monitoring of attention and arousal \cite{bhaskaran_immersive_2024}.
In contrast, bionic vision remains largely open-loop, rarely exploiting residual brain signals.
This is a critical gap. Neural interfaces are uniquely positioned to support closed-loop, co-adaptive behavior \cite{grani_improving_2024}.

Emerging work shows that pupil size, motor patterns, and noninvasive brain signals (EEG, fNIRS) can track user intent and cognitive state \cite{ding_multimodal_2019,stangl_mobile_2023}.
However, robust cognitive decoding under naturalistic conditions remains an open challenge \cite{herold_functional_2017,kvist_using_2023}.

\subsection{Output Modalities}

Most current implants deliver sparse, flickering visual percepts. 
XR systems, by contrast, routinely augment experience with haptics and spatial audio-modalities that have proven effective in enhancing spatial awareness and reducing cognitive load \cite{martin_multimodality_2021}. 

Prosthetic systems could benefit from similar multimodal strategies, as demonstrated in both accessibility tools \cite{khoo_multimodal_2016} and early XR usability studies \cite{gabbard_usability_2008}. 
Carefully designed cross-modal cues (especially when synchronized and context-aware) can reinforce key information, promote trust, and improve the user experience.

\section{The Neuroadaptive XR Design Space}

Table~\ref{tab:pipeline} outlines a new design space for neuroadaptive XR, departing from traditional visual computing and embracing the constraints of bionic vision.
Unlike conventional XR, which augments intact sensory input, neuroadaptive XR co-constructs perception with the brain through sparse, noisy, and idiosyncratic neural interfaces.

This shift demands new architectures.
A closed-loop bionic vision system (Fig.~\ref{fig:teaser}) could comprise the following components:
\begin{itemize}[leftmargin=12pt,parsep=0pt,topsep=0pt,itemsep=1pt]
    \item \textbf{Capture:} Visual input (e.g., RGB-D, thermal, egocentric video) and physiological signals (e.g., fNIRS, pupil size, IMU) are captured via lightweight, body-mounted sensors. These streams reflect both external scene dynamics and internal user state.
    
    \item \textbf{Decoding:}  A scene decoder extracts structured representations such as object identity, layout geometry, or optic flow. A user-state decoder estimates latent cognitive variables (e.g., task engagement, mental effort, arousal) from passively acquired biosignals \cite{stangl_mobile_2023,ding_multimodal_2019}. Robust multimodal fusion remains an active challenge, especially during mobile, real-world use.
    
    \item \textbf{Relevance Filtering:} This module integrates decoded scene features and user state to prioritize task-relevant information, functioning as a dynamic relevance filter. Initial implementations may employ rule-based finite-state policies (e.g., for goal-directed navigation or reading), gradually advancing toward conservative contextual bandits with offline pretraining and human-in-the-loop policy refinement \cite{beyeler_towards_2022,davari_towards_2024}.
    
    \item \textbf{Deep Stimulus Encoder (DSE):} Prioritized visual content is transformed into spatiotemporal stimulation patterns tailored to the user's implant and perceptual map \cite{granley_human---loop_2023}. Given phosphene idiosyncrasies, fading, and neural plasticity, the DSE must be coupled with: (i) per-electrode biophysical calibration under safety-constrained exploration; (ii) periodic retinotopic remapping to account for shifts in perceptual layout; and (iii) online Bayesian adaptation to update encoder parameters while enforcing hard safety constraints. These outputs drive neural stimulation, closing the loop via behavioral and neural feedback.
\end{itemize}
\vspace{0.1em}
\noindent This reframing prioritizes perceptual utility over visual fidelity. It supports interfaces that adapt to user intent, respond to environmental context, and operate through closed-loop interaction between brain, body, and world. The goal is not to replicate natural vision, but to enable actionable perception tuned to what matters.

Evaluating such systems calls for task-level, embodied benchmarks rather than pixel-level metrics. Examples include:
\begin{itemize}[leftmargin=12pt,parsep=0pt,topsep=0pt,itemsep=1pt]
\item \textbf{Navigation}: Wayfinding with static or dynamic obstacles. Metrics: success rate, time to goal, path efficiency, collisions.

\item \textbf{Object search}: Finding target items (e.g., keys) on a cluttered tabletop. Metrics: success, search time, scan redundancy.

\item \textbf{Social interaction}: Locating a speaking partner in a group. Metrics: response time, accuracy, conversational alignment.
\end{itemize}
Subjective measures (e.g., trust, agency, workload) offer complementary insights.
Early studies can begin with simulation and Wizard-of-Oz methods \cite{kasowski_immersive_2022,rasla_relative_2022}, progressing toward real-world use and co-design with blind users \cite{nadolskis_aligning_2024}.

\section{Research Questions \& Provocations}

The shift from conventional displays to direct neural stimulation introduces a new class of design, safety, and ethical challenges. Traditional \ac{XR} metrics and design principles no longer apply when output is mediated through sparse, noisy percepts, and user input is decoded from residual motor signals or neural telemetry. 

We present a set of open research questions grounded in the framework outlined in Table~\ref{tab:pipeline} to provoke discussion and shape future work in neuroadaptive XR. 
These questions are meant to guide system designers, neuroscientists, ethicists, and clinicians alike, as we navigate the emerging intersection of brain-aware computing and assistive technology.

\subsection{System Design and Representation}

\begin{enumerate}[label=\textbf{R\arabic*},leftmargin=16pt,parsep=0pt,topsep=0pt,itemsep=1pt]
  \item \textbf{Representation}: What abstractions (e.g., symbolic overlays, multimodal encodings) \cite{beyeler_towards_2022} best support scene understanding through sparse neural stimulation?
  
  \item \textbf{Multimodal Integration}: How can we coordinate audio, haptics, and phosphenes \cite{beyeler_towards_2022} to create perceptually coherent, cognitively efficient experiences?
  
  \item \textbf{Personalization}: What components of the system (e.g., encoder mappings, feedback modalities, adaptation strategies) should be personalized \cite{granley_human---loop_2023}, and how do we do so efficiently?

  \item \textbf{Continual Learning}: How do we enable long-term adaptation in both system and user \cite{eckmiller_learning_2009} while maintaining safety, stability, and interpretability?
\end{enumerate}

\subsection{Evaluation and Real-World Deployment}

\begin{enumerate}[resume, label=\textbf{R\arabic*},leftmargin=16pt,parsep=0pt,topsep=0pt,itemsep=1pt]
  \item \textbf{Evaluation}: How do we evaluate usefulness when visual quality is subjective and task-dependent \cite{peli_testing_2020}? Should we move toward embodied, task-level benchmarks (e.g., wayfinding, reading) over clinical outcomes and acuity scores?

  \item \textbf{Deployment in the Wild}: How can we bridge the lab-field gap \cite{nadolskis_aligning_2024}, particularly when deploying custom hardware with blind users in uncontrolled environments?
\end{enumerate}

\subsection{Safety, Ethics, and Societal Impact}

\begin{enumerate}[resume, label=\textbf{R\arabic*},leftmargin=16pt,parsep=0pt,topsep=0pt,itemsep=1pt]
  \item \textbf{Safety and Trust}: How can adaptive systems operating in safety-critical contexts prevent overstimulation, perceptual confusion, or failure to alert?

  \item \textbf{Neuroethics and Privacy}: Prosthetic telemetry exposes aspects of intent, attention, and emotional state \cite{goering_recommendations_2021}. Who owns this data, and how should it be protected? What norms and safeguards are needed when these systems mediate not just perception, but agency?
\end{enumerate}

\section{Conclusion}

Reframing bionic vision as neuroadaptive XR shifts the goal from restoring natural sight to enabling adaptive, task-relevant perception.
This perspective anchors bionic vision within the broader landscape of next-generation XR, opening new directions in interaction design, brain-aware computing, and embodied AI.

\acknowledgments{Supported by the National Library of Medicine of the National Institutes
of Health (NIH) under Award Number DP2-LM014268. The content is solely the responsibility of the authors and does not necessarily represent the official views of the NIH.}

\bibliographystyle{abbrv-doi}

\bibliography{references}
\end{document}